\documentclass[prl,aps,amssymb,twocolumn,superscriptaddress, notitlepage, longbibliography]{revtex4-2}

\usepackage{bm,amssymb,amsmath,dsfont,mathrsfs,amstext,latexsym,physics,relsize}
\usepackage{graphicx}

\usepackage[usenames,dvipsnames]{xcolor}
\usepackage[colorlinks=true,citecolor=teal,urlcolor=teal,linkcolor=teal]{hyperref}

\makeatletter
\def\maketitle{\@author@finish
  \title@column\titleblock@produce
  \suppressfloats[t]}
\makeatother

\newcommand{\bsff}{\beta_{\rm sff}}

\usepackage{orcidlink}

\newcommand{\figpanel}[2]{\hyperref[#1]{\ref{#1}#2}}

\begin{document}


\title{Dephasing-Induced Melting of Crystalline Spectral Form Factors}
\author{Preethi Gopalakrishnan\orcidlink{0009-0000-4053-235X}}
\email{preethi.gopalakrishnan@uni.lu}\label{email1}
\affiliation{Department  of  Physics  and  Materials  Science,  University  of  Luxembourg,  L-1511  Luxembourg, Luxembourg}
\author{Adolfo~del Campo\orcidlink{0000-0003-2219-2851}} 
\email{adolfo.delcampo@uni.lu}\label{email2}
\affiliation{Department  of  Physics  and  Materials  Science,  University  of  Luxembourg,  L-1511  Luxembourg, Luxembourg}
\affiliation{Donostia International Physics Center,  E-20018 San Sebasti\'an, Spain}
\affiliation{Institute of Science and Engineering, Kanazawa University, Kakuma-machi, Kanazawa, Ishikawa,  920-1192, Japan}

\begin{abstract}
Long-range spectral order can be probed using spectral form factors, where it manifests itself as sharp Bragg peaks at integer multiples of the Heisenberg time in nearly crystalline spectra. We examine how energy dephasing in open systems affects this spectral order and how to control it. We find that Bragg peaks fade algebraically with revival order and dephasing strength, $K_{\rm Bragg}(p)\propto(\gamma p)^{-1/2}$, within an extensive intermediate regime, and persist up to the lifetime of nearest-neighbor coherences despite an extreme decoherence rate.  
Our results establish the spectral form factor as a probe of spectral order in open quantum systems and identify temporal correlations in spectral fluctuations as a means of controlling it.
\end{abstract}

\maketitle

Spectral statistics provides a powerful tool for characterizing many-body quantum systems and plays a central role in the study of integrability, its breakdown, and the onset of quantum chaos \cite{Haake,Borgonovi16,Guhr98}. A common indicator is the level spacing distribution that is expected to be Poissonian in an integrable system, while quantum chaotic systems often exhibit Wigner-Dyson statistics~\cite{BerryTabor1977,Bohigas84}. 
The spectral characterization of complex many-body systems can be conveniently performed by integrated quantities such as the spectral form factor (SFF), which is the complex Fourier transform of the two-point energy correlator~\cite{BrezinHikami1997,Cotler2017,Kos2018,Chan2018}, and can be understood as a survival probability \cite{Leviandier86,WilkieBrumer91,Alhassid1992,Prange97}. When  using the Boltzmann factor as an energy filter, the SFF can be found in terms of partition functions at complex inverse temperature. The SFF is then equivalent to the survival probability  of a coherent Gibbs state, in both closed \cite{Papadodimas15,delcampo17,TorresHerrera2017,Maldacena2003} and open systems \cite{Xu21SFF,Cornelius21,MatsoukasRoubeas23,Matsoukas23SFFSA,MatsoukasRoubeas2024,Gopalakrishnan2026}.

Although the spectral characterization of the transition from integrability to chaos has received widespread attention,  the study of complex systems with a nearly regular spectrum has been less explored \cite{delcampo17}. The latter can be captured by random matrix ensembles with a large Dyson index $\beta$ \cite{DumitriuEdelman2002}, or strongly repulsive log gases \cite{Forrester10}. As recently pointed out by Trunin and Huse \cite{TruninHuse2026}, such spectral regularity is naturally manifested by peaks that persist over long timescales in the SFF and is attenuated by the Debye-Waller factor, which reflects the effect of energy level fluctuations in a close and beautiful analogy with the attenuation of X-ray and coherent neutron scattering by thermal motion~\cite{Debye1913,Waller1923,AshcroftMermin1976}. This observation motivates the study of crystalline spectral order and its robustness to environmental noise. 

The crystalline spectral order features an exceptionally rigid organization of energy levels, with the eigenvalues exhibiting a periodic arrangement, $x_n\simeq n$ where $n=0,\ldots,d-1$. The long-range spectral order, seen in this ``picket-fence'' like structure, gives rise to coherent Bragg revivals in the spectral form factor,
\begin{equation}
K(t)= \frac{\left|Z(\bsff+it)\right|^2}{Z(\bsff)^2} =\frac{1}{Z(\bsff)^2}\left|\sum_{n=0}^{d-1}e^{-\bsff x_n-i t x_n}\right|^2,
\end{equation}
with the partition function $Z(\bsff)=\sum_n e^{-\bsff x_n}$. The exact revivals take place at $t_p =p\,t_H$~\cite{AverbukhPerelman1989,Robinett2004}, for $p\in\mathbb{Z}$ and $t_H$, the Heisenberg time~\cite{Berry1985}. In an open system, this long-range order is expected to be progressively suppressed as the coherence between energy levels becomes evanescent and is ultimately washed away~\cite{Zurek2003,Schlosshauer2019}.

This raises a natural question: How does decoherence melt the crystalline Bragg structure, and what determines the time scale of this melting?  Here, we use ``melting'' to describe the loss of coherent Bragg structure in the spectral form factor. 

In this letter, we study the melting of spectral order induced by energy dephasing. We find that the attenuation of the Bragg peaks results from two qualitatively distinct mechanisms. The intrinsic level fluctuations produce a static Debye-Waller like suppression, whereas dephasing leads to an algebraic decay.  We further show that introducing temporal correlations in spectral fluctuations can control the decay of the Bragg peaks. 

\emph{Energy dephasing dynamics in the crystalline lattice.---} We start by analyzing the stability of the crystalline SFF of a perfectly regular spectrum under open quantum dynamics. Fluctuations in time in the Hamiltonian spectrum, as well as time coarsening of the time evolution, generally induce dephasing in the energy eigenbasis. The latter also arises in the description of quantum evolution with real clocks \cite{Egusquiza99,Egusquiza03} and frequency filters, e.g, as those used in numerical and spectral data analysis \cite{Matsoukas23SFFSA}. 
We thus consider time-continuous energy dephasing described by the master equation $\dot{\rho}=-i[H,\rho]-\gamma[H,[H,\rho]]$~\cite{Milburn1991,MoyaCessa1993,Bonifacio2000,GKS1976,Lindblad1976,BreuerPetruccione2002}, where $\gamma$ denotes the dephasing rate. 
From the short-time asymptotics of the purity decay, one can estimate the decoherence rate $1/\tau_D=\gamma \Delta H^2$, which can be upper-bounded in terms of the spectral width of the Hamiltonian \cite{Chenu17} and can grow with the Hilbert space dimension, as observed in quantum chaotic systems described by random matrix Hamiltonians \cite{Xu19}. One may thus expect energy dephasing to quickly suppress spectral signatures of the SFF, as it indeed happens in chaotic systems \cite{Xu21SFF,YangXuAdC24}. Yet, it has been claimed that energy dephasing in crystalline models is delayed to time scales much longer than the Heisenberg time \cite{TruninHuse2026}. 

We work in the energy eigenbasis, $H\ket{n}=x_n\ket{n}$, where the coherences between two energy levels with separation $\omega_{nm}=x_n-x_m$ evolve as $\rho_{nm}(t)=\rho_{nm}(0)e^{-i\omega_{nm}t-\gamma t \omega_{nm}^{2}}.$ As a result, the SFF under energy dephasing is given by
\begin{equation}
K_{\bsff}(t)=\frac{1}{Z(\bsff)^2}\sum_{n,m}
e^{-\bsff(x_n+x_m)-i\,\omega_{nm}t-\gamma t\,\omega_{nm}^2}.
\label{eq:dephased_sff}
\end{equation}
Since $\bsff >0$ simply masks the long-spectral order, we consider the infinite-temperature limit $\bsff = 0$, and for a non-degenerate (e.g., crystalline) spectrum, $Z(0)=d$, where $d$ is the Hilbert space dimension. We recall that the SFF can be understood as the survival probability $\langle\psi_{\bsff} |\rho_t|\psi_{\bsff} \rangle$ between the initial coherent Gibbs state $\psi_{\bsff} =\sum_n \sqrt{p_n}|n\rangle$, with $p_n =\exp(-\bsff x_n )/Z(\bsff)$, 
and its time evolution $\rho_t$. For $\bsff=0$, one finds that $1/\tau_D=\gamma \Delta H^2=\gamma(d^2-1)/12$. It follows that systems with a spectral crystalline structure still exhibit extreme decoherence \cite{Xu19,YangXuAdC24}, in the sense that the decoherence rate is proportional to $d^2$.

We consider the simplest crystalline spectrum of unit level spacing with $x_n=n$ ($t_H=2\pi$). The SFF can then be decomposed by neighbor-orders $k=n-m$,
\begin{equation}
K_0(t)=\frac{1}{d^2}\sum_{k=-(d-1)}^{d-1}(d-|k|)\,e^{-ikt-\gamma t\,k^2}.
\end{equation}
By analogy with Bragg diffraction, the coherent realignment of phases at $t_p= 2\pi p$ can be seen as an effect of Debye-Waller factor, when the SFF is expressed as $d\,K(2\pi p)=1+e^{-2W}d$. The resulting peaks are then given by
\begin{equation}
e^{-2W}=\frac{2}{d}\sum_{k=1}^{d-1}\Big(1-\frac kd\Big)e^{-k^2/2k_\ast^2},
\qquad k_\ast=(4\pi\gamma p)^{-1/2}.
\label{eq:e2W_exact}
\end{equation}
Within this formulation, the Gaussian envelope imposes a coherence length scale $k_\ast$, that restricts the neighbor orders $k\lesssim k_\ast$ contributing to the peaks. The fate of the revivals depends on the position of $k_\ast$ relative to the nearest neighbor ($k=1$) and the farthest neighbor ($k\sim d$), resulting in three distinct regimes:
\begin{equation}
e^{-2W}\ \simeq\
\begin{cases}
\ 1, & k_\ast\gg d\ \ (\text{rigid}),\\[3pt]
\ (2\gamma p\,d^2)^{-1/2}, & 1\ll k_\ast\ll d\ \ (\text{power-law}),\\[3pt]
\ \dfrac{2}{d}\,e^{-2\pi\gamma p}, & k_\ast\ll1\ \ (\text{exponential}).
\end{cases}
\label{eq:three_regimes}
\end{equation}

The three regimes depicted in Fig.~\ref{fig:threeregimes} can be interpreted in terms of decoherence times. A coherence spanning $k$ level spacings decays on a timescale $\tau_k=(\gamma k^2)^{-1}$. The extremal pairs with $k\sim d$, which determine the onset of decoherence $\tau_D\sim(\gamma d^2)^{-1}$, decay most rapidly. On the opposite end, the longest-lived coherence, associated with the nearest-neighbor separation $k=1$, lasts up to $\tau_\ell\equiv\tau_{k=1}=1/\gamma$. In the $\gamma \to 0$ limit, this coherence lasts indefinitely, retaining the rigid crystal order. As we increase $\gamma$, together with $p$, it delimits the regimes in Eq.~(\ref{eq:three_regimes}), corresponding to $k_\ast\sim d$ ($\gamma p\sim d^{-2}$) and $k_\ast\sim1$ ($\gamma p\sim1$). The fraction of surviving coherences $k_\ast/d$ then leads to an algebraic decay in the interval $\tau_D\ll t_p\ll\tau_\ell$, while beyond $\tau_\ell$, the loss of the final coherence gives rise to an exponential suppression. 

\begin{figure}
    \centering
    \includegraphics[width=1\columnwidth]{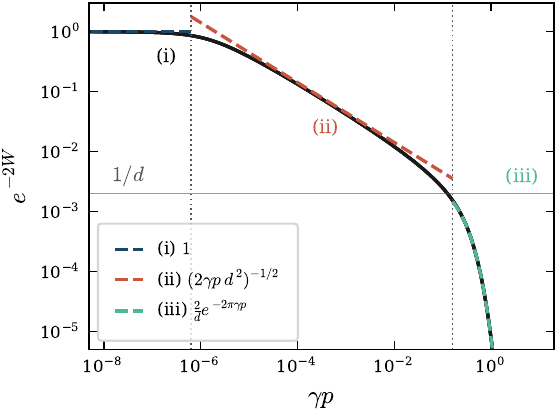}
    \caption{Three regimes of the dephasing Debye--Waller factor of a perfect crystalline lattice ($x_n=n$, $d=512$). Bragg-peak height $e^{-2W}$ as a function of $\gamma p$.}
    \label{fig:threeregimes}
\end{figure}

We note that the energy dephasing leaves the phases unchanged and instead damps the coherences $\rho_{nm}$, that carry them, at a rate $\propto \gamma k^2$. This leads to the suppression of Bragg peak heights. Nevertheless, the coherences persist up to timescales of order $1/\gamma$ that go well beyond the decoherence time $\tau_D$. 

We focus on the non-trivial intermediate regime, $d^{-2}\ll\gamma p\ll1$. In this regime, the algebraic decay results from the coherence across many neighbor orders and is therefore sensitive to the long-range order of the spectrum, which the intrinsic level fluctuations reduce. Thus, it is of interest to study crystalline behavior in the presence of competing effects simultaneously involving 
dephasing and intrinsic spectral fluctuations.  To this end, we replace the ideal picket-fence spectrum by a $\beta$-ensemble, in which the Dyson index $\beta$ provides a controlled way to tune spectral rigidity. The cases $\beta=1,2,4$ correspond to the classical Gaussian ensembles~\cite{Dyson1962c,Haake_QM_Signatures_2001,Mehta2004}, while $\beta\to\infty$ freezes the levels into the perfect rigid crystal.

\emph{Model.---} A variety of physical scenarios involving strong level repulsion are characterized by near-crystalline energy spectra, ranging from physical systems with a Coulomb gas spectrum to permutation symmetric circuits~\cite{TruninHuse2026,Forrester10,Calogero1971,Dyson1962,Sutherland1971,Bertini2025PRX,Bertini2026PRB,BogomolnyGiraudSchmit2009,BogomolnyGiraudSchmit2011,BogomolnySchmit2004}. We use the Dumitriu--Edelman tridiagonal construction which realizes the $\beta$-ensemble for arbitrary real $\beta>0$~\cite{DumitriuEdelman2002},
\begin{align}
H_\beta &= \mathrm{diag}(g_1,\ldots,g_d) +\sum_{i=1}^{d-1} \frac{a_i}{\sqrt{2}} \left(
|i\rangle\langle i+1| +\mathrm{h.c.}\right),\\
g_i &\sim \mathcal{N}(0,2),\qquad i=1,\ldots,d,\\
a_i &\sim \chi_{\beta(d-i)},\qquad i=1,\ldots,d-1.
\end{align}
The corresponding eigenvalue density is
\begin{equation}
p_\beta(\lambda)\propto\prod_{i<j}|\lambda_i-\lambda_j|^\beta\exp\left(-\frac{1}{2}\sum_i\lambda_ i^2\right),
\end{equation}
making the ensemble a convenient model for continuously tuning the degree of spectral rigidity from the classical random-matrix ensembles to the crystalline limit. The joint eigenvalue density $p_\beta(\lambda)$ can be interpreted as the Boltzmann factor of a one-dimensional Coulomb gas with charges at positions $\lambda_i$ that interact through logarithmic repulsion and are confined by a harmonic potential, where the Dyson index $\beta$ plays the role of an inverse temperature \cite{Forrester10}. In the large $\beta$ regime, the charges (levels) fluctuate weakly around their equilibrium positions $\bar{\lambda}_i$, so that $\lambda_i=\bar{\lambda}_i+\delta\lambda_i$~\cite{DumitriuEdelman2005,GorinKleptsyn2024}. After unfolding to unit mean density ($t_H = 2\pi$), $x_n=n+u_n$, the displacements are log-correlated~\cite{TruninHuse2026,DysonMehta1963,Relano2002,Faleiro2004,Riser2017},
\begin{equation}
g(k)\equiv\big\langle(u_n-u_{n+k})^2\big\rangle\simeq\frac{c}{\beta\pi^2}\ln k,
\label{eq:gk}
\end{equation}
a hallmark of the one-dimensional log-gas, which we take as our reference system, with Hamiltonian $H$ (detailed in \cite{SM}). 

\begin{figure}[t]
    \centering
    \includegraphics[width=1\columnwidth]{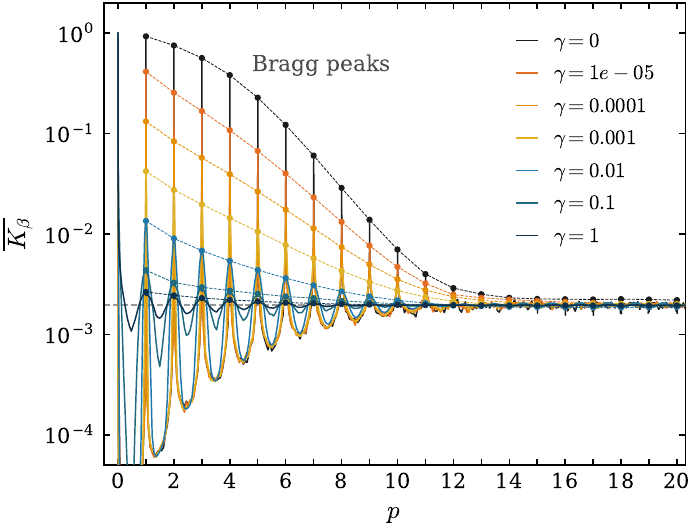}
    \caption{Melting of the crystalline order in the spectral form factor as a result of energy dephasing in DE ensembles of $d=512$ and $\beta=500$, averaged over $200$ realizations.}
    \label{fig:sff}
\end{figure}

\emph{Imperfect crystals.}--- At finite $\beta$, the residual displacements $u_n$ detune each gap $\omega_{nm}$ from $k$, and each coherence $\rho_{nm}$ acquires a phase $e^{-i2\pi p(u_n-u_m)}$ at the Bragg times. Taking the average over these displacements, we get the Bragg peaks attenuated by fluctuations, 
\begin{equation}
K_0(2\pi p)=\frac1{d^2}\sum_k(d-|k|)\,
e^{-2\pi\gamma p\,k^2}\,e^{-\frac12(2\pi p)^2 g(k)}.
\label{eq:disc_deph}
\end{equation}

The logarithmic growth of $g(k)$ then leads to algebraic suppression of the $k$-th neighbor coherence, $e^{-\frac12(2\pi p)^2g(k)}\simeq k^{-\eta_p}$, where $\eta_p=2cp^2/\beta$. The spectrum therefore exhibits quasi-long-range order, much like the algebraic Bragg peaks of two-dimensional crystals~\cite{Jancovici67,Mermin68,ImryGunther71,KosterlitzThouless1973,Peierls1935,Landau1937,Caille1972,GuntherImryLajzerowicz1980}. Bounded displacement fluctuations, by contrast, would attenuate all orders $k$ equally by the Debye--Waller factor. As the response of Bragg peaks to $\gamma$ depends on how $g(k)$ grows with the gap rather than its magnitude, tuning the dephasing strength provides a spectroscopy of spectral rigidity (see the Supplemental Material for details~\cite{SM}).

While the logarithmic growth of $g(k)$ is a universal attribute of the log-gas across all values of $\beta$~\cite{Mehta2004,ORourke2010, Gustavsson2005}, in the classical ensembles, $\beta = 1,2,4$, the Gaussian estimate leaves $\eta_p$ of order unity at the first revival. As a result, only the first few neighbor orders contribute, and the Bragg peaks dissolve into the $1/d$ plateau as expected. In these ensembles, where the SFF displays the familiar dip, ramp, and plateau, dephasing instead delays the dip and shortens the ramp~\cite{XuChenu2021}.

Figure~\ref{fig:sff} illustrates the loss of crystalline order due to intrinsic fluctuations and energy dephasing. We observe that the peaks are already suppressed due to finite-$\beta$ fluctuations. Energy dephasing further suppresses them and flattens their dependence on $p$, which nearly disappears once $k_\ast<1$ at every revival. The two mechanisms compete within a single summation, 
\begin{equation}
e^{-2W}\simeq\frac2d\sum_{k\ge1}k^{-\eta_p}\,e^{-k^2/2k_\ast^2}.
\label{eq:competition}
\end{equation}
where fluctuations determine the contribution of each neighbor order, and dephasing limits their range. At low revival order, or when $\eta_p<1$, the peaks depend strongly on neighbor-order coherence within the range $k_\ast$, as the accumulated fluctuations only weakly detune the pairs.  Dephasing thus predominantly controls the peak height, though more gradually than in the perfect crystalline spectra. The long-range coherences it removes already have partial phase spreading. Because $\eta_p$ grows quadratically with $p$, the later revivals ($\eta_p>1$) lose their long-range contributions to fluctuations and become nearly insensitive to dephasing. Dephasing now affects only the nearest neighbors, and the peak follows its decay, $e^{-\gamma t_p}$. The crossover occurs at the critical revival order,

\begin{equation}
p_c=\sqrt{\beta/2c},
\label{eq:pc}
\end{equation}
which is independent of $\gamma$ and $d$. Without dephasing, it corresponds to the timescale $t_\ast\propto t_H\sqrt\beta$ beyond which the Bragg peaks no longer grow with $d$~\cite{TruninHuse2026}, reflecting the analogy between the levels and a one-dimensional gas of fermions with long-range repulsion~\cite{Forrester10,Krivnov1982}.

The crossover at $p_c$ is observable in an open system only if $\tau_\ell\gg p_c\,t_H$, since decoherence eliminates all Bragg revivals after a time of order $\tau_\ell$. Below $p_c$, the intrinsic and extrinsic Debye--Waller exponents add within the power-law regime, $2W_p\simeq2W_p^\gamma+2W_p^\beta$, where $2W_p^\gamma=\tfrac12\ln(\gamma p)+\ln d+\tfrac12\ln2$ and $2W_p^\beta\simeq(cp^2/\beta)\ln(1/\gamma p)$. Subtracting the latter collapses these orders onto the pure-dephasing law as illustrated in Fig.~\ref{fig:Fig_wp}. The intrinsic exponent depends on $\gamma$ only through $k_\ast$, since dephasing sets the range of gaps over which the logarithmic $g(k)$ is effectively sampled. As $\gamma\to0$, this range extends over the entire spectrum, $k_\ast\to d$, and $2W_p^\beta$ saturates at the static value $\eta_p\ln d$.
\begin{figure}
    \centering
    \includegraphics[width=1\columnwidth]{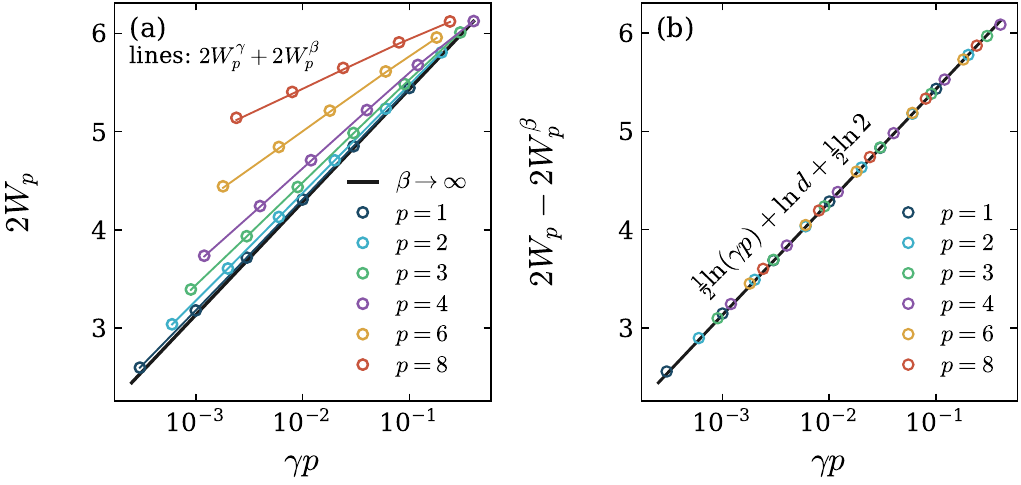}
    \caption{Dephasing-induced melting of the finite-$\beta$ spectral crystal described by the Dumitriu-Edelman ensemble with Hilbert space dimension $d=512$ and Dyson index $\beta=500$. (a) $2W_p$ versus $\gamma p$, showing the finite-$\beta$ correction to the pure-dephasing prediction. (b) Subtracting $2W_p^\beta$ restores the universal $\gamma p$ collapse onto
$2W_p^\gamma=\frac{1}{2}\ln(\gamma p)+\ln d+\frac{1}{2}\ln 2$.}
    \label{fig:Fig_wp}
\end{figure}

The interplay between intrinsic fluctuations and external decoherence can be described succinctly in terms of coherences grouped by neighbor order. The coherences on the $k$-th diagonal of $\rho$ share the dephasing rate $\gamma k^2$ to the leading order in $u_n$, while their phases retain the fluctuations $u_n-u_{n-k}$. This observation leads to the decomposition of SFF with the spectral structure factor~\cite{MartinezAzcona25},
\begin{equation}
S_k(t)\equiv\sum_n\rho_{n,n-k}(t),\qquad K(t)=\frac1d\sum_kS_k(t),
\label{eq:Sk}
\end{equation}
with initial value $S_k(0)=1-|k|/d$ at $\bsff=0$. Melting is then a loss of coherent bandwidth in $S_k$. Near each revival, $K(t_p+\delta)\simeq\frac1d\sum_kS_k(t_p)\,e^{-ik\delta}$, so this loss appears as the broadening of the Bragg peaks in time. Moreover, since intrinsic fluctuations reduce $|S_k|$ without damping the coherences, $S_k$ bounds the crystalline SFF more tightly than the quantum coherence monotone $C_{\ell_1}=\sum_{n\neq m}|\rho_{nm}|$ \cite{BaumgratzCramerPlenio2014,Streltsov2017,MarvianSpekkens2016}, which is insensitive to the phases~\cite{SM}.

\emph{Dynamical control of crystal melting.---} Drawing on the framework of motional narrowing \cite{Anderson1954,Kubo1954}, the phase spreading caused by intrinsic fluctuations can be regulated by letting the spectrum fluctuate in time, as the spreading depends on how phases accumulate. We model the fluctuations by an Ornstein--Uhlenbeck (OU) process, a stationary Gaussian process with tunable correlation time $\tau_c$~\cite{UhlenbeckOrnstein1930,Fox1986,Guha1993}. Such controlled temporal noise has been engineered in solid-state spins~\cite{deLange2010}, superconducting circuits~\cite{Potocnik2018,vonLpke2020}, trapped-ion chains~\cite{Maier2019}, and ultracold gases~\cite{Nagler2022}. We now have the instantaneous spectrum given by  $x_n(t)=n+u_n(t)$, whose fluctuations decorrelate in time as
\begin{equation}
\langle u_n(t)u_m(t')\rangle=C_{nm}e^{-|t-t'|/\tau_c},
\end{equation}
while they preserve the spectral correlations of the static crystal, $\langle[u_n(t)-u_{n-k}(t)]^2\rangle=g(k)$.
\begin{figure}
    \centering
    \includegraphics[width=1\columnwidth]{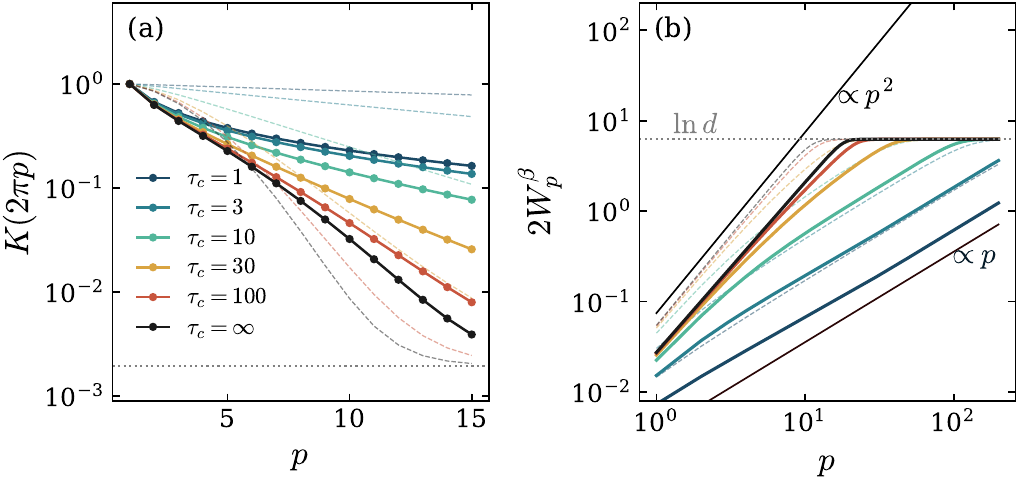}
    \caption{Dynamical control of spectral-crystal melting for $d=512$, $\beta=500$, and $\gamma=0.005$, with $\tau_c$ in units of the inverse mean level spacing. (a)~Bragg-peak heights $K(2\pi p)$ versus $t/t_H$. Faster fluctuations (smaller $\tau_c$) slow the decay relative to the static limit (black). The dotted line marks the plateau $1/d$. (b)~Debye--Waller exponent $2W_p^\beta$ versus $p$. Slow fluctuations follow the static $p^2$ scaling and fast ones grow linearly in $p$, with the change set by $p_\tau\sim\tau_c/t_H$. The dashed line shows the data for a perfect crystal, $\gamma =0$.}
    \label{fig:dynamicalcontrol}
\end{figure}
Since $H(t)$ remains diagonal in the fixed energy basis, the phase accumulated between a pair of levels is $\Phi_k(t)=\int_0^t[u_n(t')-u_{n-k}(t')]dt'$. Averaging over the fluctuations in time, we get $\langle e^{-i\Phi_k}\rangle=e^{-\frac12\mathrm{Var}\Phi_k}$~\cite{Anderson1954,Kubo1954,Kubo1969,KlauderAnderson1962}, with
\begin{equation}
\mathrm{Var}\,\Phi_k(t)=2g(k)\,\tau_c\Big[t-\tau_c\big(1-e^{-t/\tau_c}\big)\Big].
\label{eq:varphi}
\end{equation}

When we consider a long correlation time, $t\ll \tau_c$, the fluctuations have no time to evolve, leaving the spectrum essentially static, with the variance, $\mathrm{Var}\,\Phi_k\simeq g(k)t^2$. However, when $\tau_c\ll t$, the spectrum explores many configurations and the phase disorders partially cancel, resulting in the linear growth, $\mathrm{Var}\,\Phi_k\simeq2g(k)\tau_c t$. The intrinsic fluctuations then act as an additional dephasing of rate $g(k)\tau_c$, growing only as $\ln k$, while energy dephasing continues to remove coherences at rate $\gamma k^2$. In Eq.~\eqref{eq:competition}, the intrinsic fluctuation and dephasing now compete through $\eta_p^{\rm dyn}=2c\,\tau_c\,p/\pi\beta$ which is linear rather than quadratic in $p$. This shifts the critical revival order to
\begin{equation}
p_c^{\rm dyn}=\frac{\pi\beta}{2c\tau_c}.
\end{equation}
Faster fluctuations thus extend the range of revivals carried by long-range coherences, and the Bragg peaks persist to higher orders, as shown in Fig.~\ref{fig:dynamicalcontrol}. The correlation time thus provides an additional control parameter for the melting, complementary to the dephasing strength.

\emph{Conclusion and Outlook.---} We have shown the enhanced robustness of crystalline spectral form factors under energy dephasing. The Bragg peaks decay algebraically and survive up to a timescale of order $1/\gamma$ over many times the Heisenberg time, despite the extreme decoherence rate. A rich interplay between intrinsic fluctuations and energy dephasing sets a critical revival order $p_c$ that depends on spectral rigidity alone. 
Temporal correlations in the fluctuations, through motional narrowing, shift the $p_c$ and further prolong the revivals. 

The interplay between the spectra and the environment leads to both control and probing of the spectral order. While the dephasing mechanism sets the coherence length, $k_\ast$, fluctuations set the critical revival order $p_c$, encoding both environmental noise and spectral rigidity in a single SFF profile. With SFFs now accessible in quantum simulators~\cite{Joshi22,Dong25}, these findings motivate further exploration of crystalline spectral correlations under controlled noise, especially across the transition from chaotic to rigid spectra and in non-Markovian environments.

{\it Acknowledgements.--- }
The authors thank Matteo Massaro and Andr\'as Grabarits for insightful discussions. This project was supported by the Luxembourg National Research Fund (FNR Grant No.\  C24/MS/18940482/STAOpen). AdC acknowledges the Basque Government’s BERC and the Severo Ochoa Centre of Excellence Programmes (CEX2024-001491-S, funded by MICIU/AEI/10.13039/501100011033). 

\bibliography{meltingcsff}

@misc{Gopalakrishnan2026,
      title={Tailoring Quantum Chaos With Continuous Quantum Measurements}, 
      author={Preethi Gopalakrishnan and András Grabarits and Adolfo del Campo},
      year={2026},
      eprint={2602.02663},
      archivePrefix={arXiv},
      primaryClass={quant-ph},
      url={https://arxiv.org/abs/2602.02663}, 
}

@article{YangXuAdC24,
  title = {Decoherence rate in random Lindblad dynamics},
  author = {Yang, Yifeng and Xu, Zhenyu and del Campo, Adolfo},
  journal = {Phys. Rev. Res.},
  volume = {6},
  issue = {2},
  pages = {023229},
  numpages = {11},
  year = {2024},
  month = {Jun},
  publisher = {American Physical Society},
  doi = {10.1103/PhysRevResearch.6.023229},
  url = {https://link.aps.org/doi/10.1103/PhysRevResearch.6.023229}
}

@article{TruninHuse2026,
  title = {Crystalline Spectral Form Factors},
  author = {Trunin, Dmitrii A. and Huse, David A.},
  journal = {Phys. Rev. Lett.},
  volume = {136},
  issue = {25},
  pages = {250405},
  numpages = {8},
  year = {2026},
  month = {Jun},
  publisher = {American Physical Society},
  doi = {10.1103/94cc-zdbl},
  url = {https://link.aps.org/doi/10.1103/94cc-zdbl}
}

@article{TorresHerrera2017,
    author = {Torres-Herrera, E. J. and Santos, Lea F.},
    title = {Dynamical manifestations of quantum chaos: correlation hole and bulge},
    journal = {Philosophical Transactions of the Royal Society A: Mathematical, Physical and Engineering Sciences},
    volume = {375},
    number = {2108},
    pages = {20160434},
    year = {2017},
    month = {10},
    issn = {1364-503X},
    doi = {10.1098/rsta.2016.0434},
    url = {https://doi.org/10.1098/rsta.2016.0434}
}

@article{Matsoukas23SFFSA,
  title = {Unitarity breaking in self-averaging spectral form factors},
  author = {Matsoukas-Roubeas, Apollonas S. and Beau, Mathieu and Santos, Lea F. and del Campo, Adolfo},
  journal = {Phys. Rev. A},
  volume = {108},
  issue = {6},
  pages = {062201},
  numpages = {11},
  year = {2023},
  month = {Dec},
  publisher = {American Physical Society},
  doi = {10.1103/PhysRevA.108.062201},
  url = {https://link.aps.org/doi/10.1103/PhysRevA.108.062201}
}

@article{Joshi22,
  title = {Probing Many-Body Quantum Chaos with Quantum Simulators},
  author = {Joshi, Lata Kh and Elben, Andreas and Vikram, Amit and Vermersch, Beno\^{\i}t and Galitski, Victor and Zoller, Peter},
  journal = {Phys. Rev. X},
  volume = {12},
  issue = {1},
  pages = {011018},
  numpages = {34},
  year = {2022},
  month = {Jan},
  publisher = {American Physical Society},
  doi = {10.1103/PhysRevX.12.011018},
  url = {https://link.aps.org/doi/10.1103/PhysRevX.12.011018}
}

@article{Dong25,
  title = {Measuring the Spectral Form Factor in Many-Body Chaotic and Localized Phases of Quantum Processors},
  author = {Dong, Hang and Zhang, Pengfei and Da\ifmmode \breve{g}\else \u{g}\fi{}, Ceren B. and Gao, Yu and Wang, Ning and Deng, Jinfeng and Zhang, Xu and Chen, Jiachen and Xu, Shibo and Wang, Ke and Wu, Yaozu and Zhang, Chuanyu and Jin, Feitong and Zhu, Xuhao and Zhang, Aosai and Zou, Yiren and Tan, Ziqi and Cui, Zhengyi and Zhu, Zitian and Shen, Fanhao and Li, Tingting and Zhong, Jiarun and Bao, Zehang and Li, Hekang and Wang, Zhen and Guo, Qiujiang and Song, Chao and Liu, Fangli and Chan, Amos and Ying, Lei and Wang, H.},
  journal = {Phys. Rev. Lett.},
  volume = {134},
  issue = {1},
  pages = {010402},
  numpages = {7},
  year = {2025},
  month = {Jan},
  publisher = {American Physical Society},
  doi = {10.1103/PhysRevLett.134.010402},
  url = {https://link.aps.org/doi/10.1103/PhysRevLett.134.010402}
}

@article{Alhassid1992,
  title = {Spectral autocorrelation function in the statistical theory of energy levels},
  author = {Alhassid, Y. and Levine, R. D.},
  journal = {Phys. Rev. A},
  volume = {46},
  issue = {8},
  pages = {4650--4653},
  numpages = {0},
  year = {1992},
  month = {Oct},
  publisher = {American Physical Society},
  doi = {10.1103/PhysRevA.46.4650},
  url = {http://link.aps.org/doi/10.1103/PhysRevA.46.4650}
}

@Article{MatsoukasRoubeas23,
author={Matsoukas-Roubeas, Apollonas S. and Roccati, Federico and Cornelius, Julien and Xu, Zhenyu and Chenu, Aur{\'e}lia and del Campo, Adolfo},
title={Non-Hermitian Hamiltonian deformations in quantum mechanics},
journal={JHEP},
year={2023},
month={Jan},
day={13},
volume={2023},
number={1},
pages={60},
issn={1029-8479},
doi={10.1007/JHEP01(2023)060},
url={https://doi.org/10.1007/JHEP01(2023)060}
}

@misc{SM,
 note={See Supplemental Material for details}
}

@article{Prange97,
  title = {The Spectral Form Factor Is Not Self-Averaging},
  author = {Prange, R. E.},
  journal = {Phys. Rev. Lett.},
  volume = {78},
  issue = {12},
  pages = {2280--2283},
  numpages = {0},
  year = {1997},
  month = {Mar},
  publisher = {American Physical Society},
  doi = {10.1103/PhysRevLett.78.2280},
  url = {https://link.aps.org/doi/10.1103/PhysRevLett.78.2280}
}

@article{Papadodimas15,
  title = {Local Operators in the Eternal Black Hole},
  author = {Papadodimas, Kyriakos and Raju, Suvrat},
  journal = {Phys. Rev. Lett.},
  volume = {115},
  issue = {21},
  pages = {211601},
  numpages = {6},
  year = {2015},
  month = {Nov},
  publisher = {American Physical Society},
  doi = {10.1103/PhysRevLett.115.211601},
  url = {https://link.aps.org/doi/10.1103/PhysRevLett.115.211601}
}

@article{Guhr98,
title = "Random-matrix theories in quantum physics: common concepts",
journal = "Phys. Rep",
volume = "299",
number = "4",
pages = "189 - 425",
year = "1998",
issn = "0370-1573",
doi = "https://doi.org/10.1016/S0370-1573(97)00088-4",
url = "http://www.sciencedirect.com/science/article/pii/S0370157397000884",
author = {Thomas Guhr and Axel M\"uller-Groeling and Hans A. Weidenm\"uller}
}

@article{Bohigas84,
  title = {Characterization of Chaotic Quantum Spectra and Universality of Level Fluctuation Laws},
  author = {Bohigas, O. and Giannoni, M. J. and Schmit, C.},
  journal = {Phys. Rev. Lett.},
  volume = {52},
  issue = {1},
  pages = {1--4},
  numpages = {0},
  year = {1984},
  month = {Jan},
  publisher = {American Physical Society},
  doi = {10.1103/PhysRevLett.52.1},
  url = {https://link.aps.org/doi/10.1103/PhysRevLett.52.1}
}

@article{MartinezAzcona25,
  title = {Decomposing the spectral form factor},
  author = {Martinez-Azcona, Pablo and Shir, Ruth and Chenu, Aur\'elia},
  journal = {Phys. Rev. B},
  volume = {111},
  issue = {16},
  pages = {165108},
  numpages = {17},
  year = {2025},
  month = {Apr},
  publisher = {American Physical Society},
  doi = {10.1103/PhysRevB.111.165108},
  url = {https://link.aps.org/doi/10.1103/PhysRevB.111.165108}
}

@article{Borgonovi16,
title = {Quantum chaos and thermalization in isolated systems of interacting particles},
journal = {Phys. Rep},
volume = {626},
pages = {1-58},
year = {2016},
note = {Quantum chaos and thermalization in isolated systems of interacting particles},
issn = {0370-1573},
doi = {https://doi.org/10.1016/j.physrep.2016.02.005},
url = {https://www.sciencedirect.com/science/article/pii/S0370157316000831},
author = {F. Borgonovi and F.M. Izrailev and L.F. Santos and V.G. Zelevinsky}
}

@article{Xu21SFF,
  title = {Thermofield dynamics: Quantum chaos versus decoherence},
  author = {Xu, Z. and Chenu, A. and Prosen, T. and del Campo, A.},
  journal = {Phys. Rev. B},
  volume = {103},
  issue = {6},
  pages = {064309},
  numpages = {11},
  year = {2021},
  month = {Feb},
  publisher = {American Physical Society},
  doi = {10.1103/PhysRevB.103.064309},
  url = {https://link.aps.org/doi/10.1103/PhysRevB.103.064309}
}

@article{Cornelius21,
  title = {Spectral Filtering Induced by Non-Hermitian Evolution with Balanced Gain and Loss: Enhancing Quantum Chaos},
  author = {Cornelius, Julien and Xu, Zhenyu and Saxena, Avadh and Chenu, Aur\'elia and del Campo, Adolfo},
  journal = {Phys. Rev. Lett.},
  volume = {128},
  issue = {19},
  pages = {190402},
  numpages = {6},
  year = {2022},
  month = {May},
  publisher = {American Physical Society},
  doi = {10.1103/PhysRevLett.128.190402},
  url = {https://link.aps.org/doi/10.1103/PhysRevLett.128.190402}
}

@article{Leviandier86,
  title = {Fourier Transform: A Tool to Measure Statistical Level Properties in Very Complex Spectra},
  author = {Leviandier, Luc and Lombardi, Maurice and Jost, R\'emi and Pique, Jean Paul},
  journal = {Phys. Rev. Lett.},
  volume = {56},
  issue = {23},
  pages = {2449--2452},
  numpages = {0},
  year = {1986},
  month = {Jun},
  publisher = {American Physical Society},
  doi = {10.1103/PhysRevLett.56.2449},
  url = {https://link.aps.org/doi/10.1103/PhysRevLett.56.2449}
}

@article{WilkieBrumer91,
  title = {Time-dependent manifestations of quantum chaos},
  author = {Wilkie, Joshua and Brumer, Paul},
  journal = {Phys. Rev. Lett.},
  volume = {67},
  issue = {10},
  pages = {1185--1188},
  numpages = {0},
  year = {1991},
  month = {Sep},
  publisher = {American Physical Society},
  doi = {10.1103/PhysRevLett.67.1185},
  url = {https://link.aps.org/doi/10.1103/PhysRevLett.67.1185}
}

@book{Haake,
  title = {Quantum Signatures of Chaos},
  publisher = {Springer},
  year = {2010},
  author = {Haake, F.},
  address = {Berlin},
  url = {https://doi.org/10.1007/978-3-642-05428-0}
}

@book{Forrester10,
  title={Log-Gases and Random Matrices (LMS-34)},
  author={Forrester, P.J.},
  isbn={9781400835416},
  lccn={2009053314},
  series={London Mathematical Society Monographs},
  url={https://books.google.lu/books?id=C7z3NgOlb1gC},
  year={2010},
  publisher={Princeton University Press}
}

@article{Chenu17,
  title = {Quantum Simulation of Generic Many-Body Open System Dynamics Using Classical Noise},
  author = {Chenu, A. and Beau, M. and Cao, J. and del Campo, A.},
  journal = {Phys. Rev. Lett.},
  volume = {118},
  issue = {14},
  pages = {140403},
  numpages = {6},
  year = {2017},
  month = {Apr},
  publisher = {American Physical Society},
  doi = {10.1103/PhysRevLett.118.140403},
  url = {https://link.aps.org/doi/10.1103/PhysRevLett.118.140403}
}

@article{Egusquiza03,
  title = {Real clocks and the Zeno effect},
  author = {Egusquiza, I. L. and Garay, Luis J.},
  journal = {Phys. Rev. A},
  volume = {68},
  issue = {2},
  pages = {022104},
  numpages = {10},
  year = {2003},
  month = {Aug},
  publisher = {American Physical Society},
  doi = {10.1103/PhysRevA.68.022104},
  url = {https://link.aps.org/doi/10.1103/PhysRevA.68.022104}
}

@article{Egusquiza99,
  title = {Quantum evolution according to real clocks},
  author = {Egusquiza, I. L. and Garay, L. J. and Raya, J. M.},
  journal = {Phys. Rev. A},
  volume = {59},
  issue = {5},
  pages = {3236--3240},
  numpages = {0},
  year = {1999},
  month = {May},
  publisher = {American Physical Society},
  doi = {10.1103/PhysRevA.59.3236},
  url = {https://link.aps.org/doi/10.1103/PhysRevA.59.3236}
}

@article{Xu19,
  title = {Extreme Decoherence and Quantum Chaos},
  author = {Xu, Zhenyu and Garc\'{\i}a-Pintos, Luis Pedro and Chenu, Aur\'elia and del Campo, Adolfo},
  journal = {Phys. Rev. Lett.},
  volume = {122},
  issue = {1},
  pages = {014103},
  numpages = {6},
  year = {2019},
  month = {Jan},
  publisher = {American Physical Society},
  doi = {10.1103/PhysRevLett.122.014103},
  url = {https://link.aps.org/doi/10.1103/PhysRevLett.122.014103}
}

@article{delcampo17,
  title = {Scrambling the spectral form factor: Unitarity constraints and exact results},
  author = {del Campo, A. and Molina-Vilaplana, J. and Sonner, J.},
  journal = {Phys. Rev. D},
  volume = {95},
  issue = {12},
  pages = {126008},
  numpages = {16},
  year = {2017},
  month = {Jun},
  publisher = {American Physical Society},
  doi = {10.1103/PhysRevD.95.126008},
  url = {https://link.aps.org/doi/10.1103/PhysRevD.95.126008}
}

@book{Haake_QM_Signatures_2001,
author = {Haake, Fritz},
year = {2001},
month = {01},
pages = {},
title = {Quantum Signatures of Chaos},
volume = {54},
isbn = {978-3-642-08722-6},
doi = {10.1007/978-3-662-04506-0}
}

@article{UhlenbeckOrnstein1930,
  author = {G. E. Uhlenbeck and L. S. Ornstein},
  title = {On the Theory of the {Brownian} Motion},
  journal = {Phys. Rev.},
  volume = {36},
  number = {5},
  pages = {823--841},
  year = {1930},
  doi = {10.1103/PhysRev.36.823}
}

@article{Anderson1954,
  author = {P. W. Anderson},
  title = {A Mathematical Model for the Narrowing of Spectral Lines by Exchange or Motion},
  journal = {J. Phys. Soc. Jpn.},
  volume = {9},
  number = {3},
  pages = {316--339},
  year = {1954},
  doi = {10.1143/JPSJ.9.316}
}

@article{Guha1993,
  author = {S. Guha and G. Sanyal and S. H. Mandal and D. Mukherjee},
  title = {Stochastic averaging of the time-evolution operator for quantum systems driven by {Ornstein-Uhlenbeck} colored noise: A nonperturbative cluster cumulant method},
  journal = {Phys. Rev. E},
  volume = {47},
  number = {4},
  pages = {2336--2356},
  year = {1993},
  doi = {10.1103/PhysRevE.47.2336}
}

@article{Cotler2017,
  author  = {Cotler, Jordan S. and Gur-Ari, Guy and Hanada, Masanori and
             Polchinski, Joseph and Saad, Phil and Shenker, Stephen H. and
             Stanford, Douglas and Streicher, Alexandre and Tezuka, Masaki},
  title   = {Black Holes and Random Matrices},
  journal = {J. High Energy Phys.},
  volume  = {2017},
  number  = {05},
  pages   = {118},
  year    = {2017},
  doi     = {10.1007/JHEP05(2017)118},
  note    = {Erratum: J. High Energy Phys. \\textbf{2018}, 09, 002 (2018)},
  archivePrefix = {arXiv},
  eprint  = {1611.04650},
}

@article{DumitriuEdelman2002,
  author  = {Dumitriu, Ioana and Edelman, Alan},
  title   = {Matrix Models for Beta Ensembles},
  journal = {J. Math. Phys.},
  volume  = {43},
  number  = {11},
  pages   = {5830--5847},
  year    = {2002},
  doi     = {10.1063/1.1507823},
  archivePrefix = {arXiv},
  eprint  = {math-ph/0206043},
}

@article{XuChenu2021,
  author  = {Xu, Zhenyu and Chenu, Aur\'elia and Prosen, Toma\v{z} and
             del Campo, Adolfo},
  title   = {Thermofield Dynamics: Quantum Chaos versus Decoherence},
  journal = {Phys. Rev. B},
  volume  = {103},
  pages   = {064309},
  year    = {2021},
  doi     = {10.1103/PhysRevB.103.064309},
}

@article{Dyson1962c,
  author  = {Dyson, Freeman J.},
  title   = {The Threefold Way. {A}lgebraic Structure of Symmetry Groups and Ensembles in Quantum Mechanics},
  journal = {Journal of Mathematical Physics},
  volume  = {3},
  number  = {6},
  pages   = {1199--1215},
  year    = {1962}
}

@book{Mehta2004,
  author    = {Mehta, Madan Lal},
  title     = {Random Matrices},
  edition   = {3rd},
  publisher = {Elsevier/Academic Press},
  address   = {Amsterdam},
  year      = {2004},
}

@article{Lindblad1976,
  author  = {Lindblad, G\"oran},
  title   = {On the Generators of Quantum Dynamical Semigroups},
  journal = {Commun. Math. Phys.},
  volume  = {48},
  pages   = {119},
  year    = {1976},
  doi     = {10.1007/BF01608499},
}

@article{GKS1976,
  author  = {Gorini, Vittorio and Kossakowski, Andrzej and
             Sudarshan, E. C. G.},
  title   = {Completely Positive Dynamical Semigroups of $N$-Level Systems},
  journal = {J. Math. Phys.},
  volume  = {17},
  pages   = {821},
  year    = {1976},
  doi     = {10.1063/1.522979},
}

@article{BaumgratzCramerPlenio2014,
  title = {Quantifying Coherence},
  author = {Baumgratz, T. and Cramer, M. and Plenio, M. B.},
  journal = {Phys. Rev. Lett.},
  volume = {113},
  issue = {14},
  pages = {140401},
  numpages = {5},
  year = {2014},
  month = {Sep},
  publisher = {American Physical Society},
  doi = {10.1103/PhysRevLett.113.140401},
  url = {https://link.aps.org/doi/10.1103/PhysRevLett.113.140401}
}

@article{MatsoukasRoubeas2024,
  title   = {Quantum Chaos and Coherence: Random Parametric Quantum Channels},
  author  = {Matsoukas-Roubeas, Apollonas S. and Prosen, Toma{\v z} and del Campo, Adolfo},
  journal = {Quantum},
  volume  = {8},
  pages   = {1446},
  year    = {2024},
  doi     = {10.22331/q-2024-08-27-1446}
}

@article{Fox1986,
  title = {Stochastic theory of line shape and relaxation},
  author = {Faid, Karim and Fox, Ronald F.},
  journal = {Phys. Rev. A},
  volume = {34},
  issue = {5},
  pages = {4286--4302},
  numpages = {0},
  year = {1986},
  month = {Nov},
  publisher = {American Physical Society},
  doi = {10.1103/PhysRevA.34.4286},
  url = {https://link.aps.org/doi/10.1103/PhysRevA.34.4286}
}

@article{deLange2010,
  title = {Universal Dynamical Decoupling of a Single Solid-State Spin from a Spin Bath},
  volume = {330},
  ISSN = {1095-9203},
  url = {http://dx.doi.org/10.1126/science.1192739},
  DOI = {10.1126/science.1192739},
  number = {6000},
  journal = {Science},
  publisher = {American Association for the Advancement of Science (AAAS)},
  author = {de Lange,  G. and Wang,  Z. H. and Ristè,  D. and Dobrovitski,  V. V. and Hanson,  R.},
  year = {2010},
  month = Oct,
  pages = {60–63}
}

@article{Potocnik2018,
  author    = {Poto{\v{c}}nik, Anton and Bargerbos, Arno and Schr{\"o}der, Florian A. Y. N. and Khan, Saeed A. and Collodo, Michele C. and Gasparinetti, Simone and Salath{\'e}, Yves and Creatore, Celestino and Eichler, Christopher and T{\"u}reci, Hakan E. and Chin, Alex W. and Wallraff, Andreas},
  title     = {Studying light-harvesting models with superconducting circuits},
  journal   = {Nat. Commun.},
  volume    = {9},
  number    = {1},
  pages     = {904},
  year      = {2018},
  doi       = {10.1038/s41467-018-03312-x}
}

@article{Maier2019,
  title = {Environment-Assisted Quantum Transport in a 10-qubit Network},
  author = {Maier, Christine and Brydges, Tiff and Jurcevic, Petar and Trautmann, Nils and Hempel, Cornelius and Lanyon, Ben P. and Hauke, Philipp and Blatt, Rainer and Roos, Christian F.},
  journal = {Phys. Rev. Lett.},
  volume = {122},
  issue = {5},
  pages = {050501},
  numpages = {6},
  year = {2019},
  month = {Feb},
  publisher = {American Physical Society},
  doi = {10.1103/PhysRevLett.122.050501},
  url = {https://link.aps.org/doi/10.1103/PhysRevLett.122.050501}
}

@article{Nagler2022,
  title = {Ultracold Bose Gases in Dynamic Disorder with Tunable Correlation Time},
  author = {Nagler, Benjamin and Will, Martin and Hiebel, Silvia and Barbosa, Sian and Koch, Jennifer and Fleischhauer, Michael and Widera, Artur},
  journal = {Phys. Rev. Lett.},
  volume = {128},
  issue = {23},
  pages = {233601},
  numpages = {6},
  year = {2022},
  month = {Jun},
  publisher = {American Physical Society},
  doi = {10.1103/PhysRevLett.128.233601},
  url = {https://link.aps.org/doi/10.1103/PhysRevLett.128.233601}
}

@article{vonLpke2020,
  title = {Two-Qubit Spectroscopy of Spatiotemporally Correlated Quantum Noise in Superconducting Qubits},
  author = {von L{\"u}pke, Uwe and Beaudoin, F{\'e}lix and Norris, Leigh M. and Sung, Youngkyu and Winik, Roni and Qiu, Jack Y. and Kjaergaard, Morten and Kim, David and Yoder, Jonilyn and Gustavsson, Simon and Viola, Lorenza and Oliver, William D.},
  journal = {PRX Quantum},
  volume = {1},
  issue = {1},
  pages = {010305},
  numpages = {23},
  year = {2020},
  month = {Sep},
  publisher = {American Physical Society},
  doi = {10.1103/PRXQuantum.1.010305},
  url = {https://link.aps.org/doi/10.1103/PRXQuantum.1.010305}
}

@article{Bogomolny2009Lax,
  author  = {Bogomolny, E. and Giraud, O. and Schmit, C.},
  title   = {Random Matrix Ensembles Associated with Lax Matrices},
  journal = {Phys. Rev. Lett.},
  volume  = {103},
  pages   = {054103},
  year    = {2009},
  doi     = {10.1103/PhysRevLett.103.054103}
}

@article{Bogomolny2011,
  author        = {Bogomolny, E. and Giraud, O. and Schmit, C.},
  title         = {Integrable random matrix ensembles},
  journal       = {Nonlinearity},
  volume        = {24},
  number        = {11},
  pages         = {3179--3213},
  year          = {2011},
  publisher     = {IOP Publishing},
  doi           = {10.1088/0951-7715/24/11/010}
}

@article{Dyson1962,
  author    = {Dyson, Freeman J.},
  title     = {Statistical Theory of the Energy Levels of Complex Systems. {I}},
  journal   = {J. Math. Phys.},
  volume    = {3},
  number    = {1},
  pages     = {140--156},
  year      = {1962},
  doi       = {10.1063/1.1703773}
}

@article{Sutherland1971,
  author    = {Sutherland, Bill},
  title     = {Quantum Many-Body Problem in One Dimension: Ground State},
  journal   = {J. Math. Phys.},
  volume    = {12},
  number    = {2},
  pages     = {246--250},
  year      = {1971},
  doi       = {10.1063/1.1665584}
}

@article{Calogero1971,
  author    = {Calogero, Francesco},
  title     = {Solution of the One-Dimensional {$N$}-Body Problems with Quadratic and/or Inversely Quadratic Pair Potentials},
  journal   = {J. Math. Phys.},
  volume    = {12},
  number    = {3},
  pages     = {419--436},
  year      = {1971},
  doi       = {10.1063/1.1665604}
}

@article{Jancovici67,
  author    = {Jancovici, B.},
  title     = {Infinite Susceptibility Without Long-Range Order: The Two-Dimensional Harmonic ``Solid''},
  journal   = {Phys. Rev. Lett.},
  volume    = {19},
  number    = {1},
  pages     = {20--22},
  year      = {1967},
  doi       = {10.1103/PhysRevLett.19.20}
}

@article{Mermin68,
  title = {Crystalline Order in Two Dimensions},
  volume = {176},
  ISSN = {0031-899X},
  url = {http://dx.doi.org/10.1103/PhysRev.176.250},
  DOI = {10.1103/physrev.176.250},
  number = {1},
  journal = {Physical Review},
  publisher = {American Physical Society (APS)},
  author = {Mermin,  N. D.},
  year = {1968},
  month = Dec,
  pages = {250–254}
}

@article{ImryGunther71,
  author    = {Imry, Y. and Gunther, L.},
  title     = {Fluctuations and Physical Properties of the Two-Dimensional Crystal Lattice},
  journal   = {Phys. Rev. B},
  volume    = {3},
  number    = {11},
  pages     = {3939--3945},
  year      = {1971},
  doi       = {10.1103/PhysRevB.3.3939}
}

@article{Milburn1991,
  title   = {Intrinsic decoherence in quantum mechanics},
  author  = {Milburn, G. J.},
  journal = {Phys. Rev. A},
  volume  = {44},
  number  = {9},
  pages   = {5401--5406},
  year    = {1991},
  doi     = {10.1103/PhysRevA.44.5401}
}

@article{MoyaCessa1993,
  title   = {Intrinsic decoherence in the atom-field interaction},
  author  = {Moya-Cessa, H. and Bu{\v{z}}ek, V. and Kim, M. S. and Knight, P. L.},
  journal = {Phys. Rev. A},
  volume  = {48},
  number  = {5},
  pages   = {3900--3905},
  year    = {1993},
  doi     = {10.1103/PhysRevA.48.3900}
}

@article{Bonifacio2000,
  title   = {Model-independent approach to nondissipative decoherence},
  author  = {Bonifacio, Rodolfo and Olivares, Stefano and Tombesi, Paolo and Vitali, David},
  journal = {Phys. Rev. A},
  volume  = {61},
  number  = {5},
  pages   = {053802},
  year    = {2000},
  doi     = {10.1103/PhysRevA.61.053802}
}

@article{Robinett2004,
  title   = {Quantum wave packet revivals},
  author  = {Robinett, R. W.},
  journal = {Phys. Rep.},
  volume  = {392},
  number  = {1--2},
  pages   = {1--119},
  year    = {2004},
  doi     = {10.1016/j.physrep.2003.11.002}
}

@article{AverbukhPerelman1989,
  title   = {Fractional revivals: Universality in the long-term evolution of quantum wave packets beyond the correspondence principle dynamics},
  author  = {Averbukh, I. Sh. and Perelman, N. F.},
  journal = {Phys. Lett. A},
  volume  = {139},
  number  = {9},
  pages   = {449--453},
  year    = {1989},
  doi     = {10.1016/0375-9601(89)90943-2}
}

@book{BreuerPetruccione2002,
  title     = {The Theory of Open Quantum Systems},
  author    = {Breuer, Heinz-Peter and Petruccione, Francesco},
  publisher = {Oxford University Press},
  address   = {Oxford},
  year      = {2002}
}

@article{Zurek2003,
  title   = {Decoherence, einselection, and the quantum origins of the classical},
  author  = {Zurek, Wojciech Hubert},
  journal = {Rev. Mod. Phys.},
  volume  = {75},
  number  = {3},
  pages   = {715--775},
  year    = {2003},
  doi     = {10.1103/RevModPhys.75.715}
}

@article{Schlosshauer2019,
  title   = {Quantum decoherence},
  author  = {Schlosshauer, Maximilian},
  journal = {Phys. Rep.},
  volume  = {831},
  pages   = {1--57},
  year    = {2019},
  doi     = {10.1016/j.physrep.2019.10.001}
}

@article{BerryTabor1977,
  title   = {Level clustering in the regular spectrum},
  author  = {Berry, Michael Victor and Tabor, M.},
  journal = {Proc. R. Soc. Lond. A},
  volume  = {356},
  number  = {1686},
  pages   = {375--394},
  year    = {1977},
  doi     = {10.1098/rspa.1977.0140}
}

@article{Berry1985,
  title   = {Semiclassical theory of spectral rigidity},
  author  = {Berry, Michael Victor},
  journal = {Proc. R. Soc. Lond. A},
  volume  = {400},
  number  = {1819},
  pages   = {229--251},
  year    = {1985},
  doi     = {10.1098/rspa.1985.0078}
}

@article{BrezinHikami1997,
  title   = {Spectral form factor in a random matrix theory},
  author  = {Br{\'e}zin, E. and Hikami, S.},
  journal = {Phys. Rev. E},
  volume  = {55},
  number  = {4},
  pages   = {4067--4083},
  year    = {1997},
  doi     = {10.1103/PhysRevE.55.4067}
}

@article{Kos2018,
  title   = {Many-body quantum chaos: Analytic connection to random matrix theory},
  author  = {Kos, Pavel and Ljubotina, Marko and Prosen, Toma{\v{z}}},
  journal = {Phys. Rev. X},
  volume  = {8},
  number  = {2},
  pages   = {021062},
  year    = {2018},
  doi     = {10.1103/PhysRevX.8.021062}
}

@article{Chan2018,
  title   = {Solution of a minimal model for many-body quantum chaos},
  author  = {Chan, Amos and De Luca, Andrea and Chalker, J. T.},
  journal = {Phys. Rev. X},
  volume  = {8},
  number  = {4},
  pages   = {041019},
  year    = {2018},
  doi     = {10.1103/PhysRevX.8.041019}
}

@article{Maldacena2003,
  title   = {Eternal black holes in anti-de Sitter},
  author  = {Maldacena, Juan},
  journal = {J. High Energy Phys.},
  volume  = {2003},
  number  = {04},
  pages   = {021},
  year    = {2003},
  doi     = {10.1088/1126-6708/2003/04/021}
}

@article{Bertini2025PRX,
  title   = {Quantum and classical dynamics with random permutation circuits},
  author  = {Bertini, Bruno and Klobas, Katja and Kos, Pavel and Malz, Daniel},
  journal = {Phys. Rev. X},
  volume  = {15},
  number  = {1},
  pages   = {011015},
  year    = {2025},
  doi     = {10.1103/PhysRevX.15.011015}
}

@article{Bertini2026PRB,
  title   = {Random permutation circuits beyond qubits are quantum chaotic},
  author  = {Bertini, Bruno and Klobas, Katja and Kos, Pavel and Malz, Daniel},
  journal = {Phys. Rev. B},
  volume  = {113},
  number  = {10},
  pages   = {L100302},
  year    = {2026},
  doi     = {10.1103/ql9x-6kzj}
}

@article{BogomolnyGiraudSchmit2009,
  title   = {Random matrix ensembles associated with {L}ax matrices},
  author  = {Bogomolny, E. and Giraud, O. and Schmit, C.},
  journal = {Phys. Rev. Lett.},
  volume  = {103},
  number  = {5},
  pages   = {054103},
  year    = {2009},
  doi     = {10.1103/PhysRevLett.103.054103}
}

@article{BogomolnyGiraudSchmit2011,
  title   = {Integrable random matrix ensembles},
  author  = {Bogomolny, E. and Giraud, O. and Schmit, C.},
  journal = {Nonlinearity},
  volume  = {24},
  number  = {11},
  pages   = {3179--3213},
  year    = {2011},
  doi     = {10.1088/0951-7715/24/11/010}
}

@article{BogomolnySchmit2004,
  title   = {Spectral statistics of a quantum interval-exchange map},
  author  = {Bogomolny, E. and Schmit, C.},
  journal = {Phys. Rev. Lett.},
  volume  = {93},
  number  = {25},
  pages   = {254102},
  year    = {2004},
  doi     = {10.1103/PhysRevLett.93.254102}
}

@article{DumitriuEdelman2005,
  title   = {Eigenvalues of {H}ermite and {L}aguerre ensembles: large beta asymptotics},
  author  = {Dumitriu, Ioana and Edelman, Alan},
  journal = {Ann. Inst. Henri Poincar{\'e} Probab. Stat.},
  volume  = {41},
  number  = {6},
  pages   = {1083--1099},
  year    = {2005},
  doi     = {10.1016/j.anihpb.2004.11.002}
}

@article{GorinKleptsyn2024,
  title   = {Universal objects of the infinite beta random matrix theory},
  author  = {Gorin, Vadim and Kleptsyn, Victor},
  journal = {J. Eur. Math. Soc.},
  volume  = {26},
  number  = {9},
  pages   = {3429--3496},
  year    = {2024},
  doi     = {10.4171/JEMS/1336}
}

@article{DysonMehta1963,
  title   = {Statistical theory of the energy levels of complex systems. {IV}},
  author  = {Dyson, Freeman J. and Mehta, Madan Lal},
  journal = {J. Math. Phys.},
  volume  = {4},
  number  = {5},
  pages   = {701--712},
  year    = {1963},
  doi     = {10.1063/1.1704008}
}

@article{Gustavsson2005,
  author  = {Gustavsson, Jonas},
  title   = {Gaussian fluctuations of eigenvalues in the {GUE}},
  journal = {Ann. Inst. H. Poincar\'e Probab. Statist.},
  volume  = {41},
  number  = {2},
  pages   = {151--178},
  year    = {2005},
  doi     = {10.1016/j.anihpb.2004.04.002}
}

@article{ORourke2010,
  author  = {O'Rourke, Sean},
  title   = {Gaussian Fluctuations of Eigenvalues in {W}igner Random Matrices},
  journal = {J. Stat. Phys.},
  volume  = {138},
  number  = {6},
  pages   = {1045--1066},
  year    = {2010},
  doi     = {10.1007/s10955-009-9906-y}
}

@article{Relano2002,
  title   = {Quantum chaos and $1/f$ noise},
  author  = {Rela{\~n}o, A. and G{\'o}mez, J. M. G. and Molina, R. A. and Retamosa, J. and Faleiro, E.},
  journal = {Phys. Rev. Lett.},
  volume  = {89},
  number  = {24},
  pages   = {244102},
  year    = {2002},
  doi     = {10.1103/PhysRevLett.89.244102}
}

@article{Faleiro2004,
  title   = {Theoretical derivation of $1/f$ noise in quantum chaos},
  author  = {Faleiro, E. and G{\'o}mez, J. M. G. and Molina, R. A. and Mu{\~n}oz, L. and Rela{\~n}o, A. and Retamosa, J.},
  journal = {Phys. Rev. Lett.},
  volume  = {93},
  number  = {24},
  pages   = {244101},
  year    = {2004},
  doi     = {10.1103/PhysRevLett.93.244101}
}

@article{Riser2017,
  title   = {Power spectrum of long eigenlevel sequences in quantum chaotic systems},
  author  = {Riser, Roman and Osipov, Vladimir Al. and Kanzieper, Eugene},
  journal = {Phys. Rev. Lett.},
  volume  = {118},
  number  = {20},
  pages   = {204101},
  year    = {2017},
  doi     = {10.1103/PhysRevLett.118.204101}
}

@book{AshcroftMermin1976,
  title     = {Solid State Physics},
  author    = {Ashcroft, Neil W. and Mermin, N. David},
  publisher = {Holt, Rinehart and Winston},
  address   = {New York},
  year      = {1976}
}

@article{Debye1913,
  title   = {Interferenz von {R}{\"o}ntgenstrahlen und {W}{\"a}rmebewegung},
  author  = {Debye, P.},
  journal = {Ann. Phys.},
  volume  = {348},
  number  = {1},
  pages   = {49--92},
  year    = {1913},
  doi     = {10.1002/andp.19133480105}
}

@article{Waller1923,
  title   = {Zur {F}rage der {E}inwirkung der {W}{\"a}rmebewegung auf die {I}nterferenz von {R}{\"o}ntgenstrahlen},
  author  = {Waller, Ivar},
  journal = {Z. Phys.},
  volume  = {17},
  number  = {1},
  pages   = {398--408},
  year    = {1923},
  doi     = {10.1007/BF01328696}
}

@article{Peierls1935,
  title   = {Quelques propri{\'e}t{\'e}s typiques des corps solides},
  author  = {Peierls, R. E.},
  journal = {Ann. Inst. Henri Poincar{\'e}},
  volume  = {5},
  pages   = {177--222},
  year    = {1935}
}

@article{Landau1937,
  title   = {Zur {T}heorie der {P}hasenumwandlungen {II}},
  author  = {Landau, L. D.},
  journal = {Phys. Z. Sowjetunion},
  volume  = {11},
  pages   = {26--35},
  year    = {1937}
}

@article{Caille1972,
  title   = {Remarques sur la diffusion des rayons {X} dans les smectiques},
  author  = {Caill{\'e}, A.},
  journal = {C. R. Acad. Sci. Paris B},
  volume  = {274},
  pages   = {891--893},
  year    = {1972}
}

@article{GuntherImryLajzerowicz1980,
  title   = {X-ray scattering in smectic-$A$ liquid crystals},
  author  = {Gunther, Leon and Imry, Yoseph and Lajzerowicz, Joseph},
  journal = {Phys. Rev. A},
  volume  = {22},
  number  = {4},
  pages   = {1733--1740},
  year    = {1980},
  doi     = {10.1103/PhysRevA.22.1733}
}

@article{KosterlitzThouless1973,
  title   = {Ordering, metastability and phase transitions in two-dimensional systems},
  author  = {Kosterlitz, J. M. and Thouless, D. J.},
  journal = {J. Phys. C: Solid State Phys.},
  volume  = {6},
  number  = {7},
  pages   = {1181--1203},
  year    = {1973},
  doi     = {10.1088/0022-3719/6/7/010}
}

@article{MarvianSpekkens2016,
  title   = {How to quantify coherence: Distinguishing speakable and unspeakable notions},
  author  = {Marvian, Iman and Spekkens, Robert W.},
  journal = {Phys. Rev. A},
  volume  = {94},
  number  = {5},
  pages   = {052324},
  year    = {2016},
  doi     = {10.1103/PhysRevA.94.052324}
}

@article{Streltsov2017,
  title   = {\textit{Colloquium}: Quantum coherence as a resource},
  author  = {Streltsov, Alexander and Adesso, Gerardo and Plenio, Martin B.},
  journal = {Rev. Mod. Phys.},
  volume  = {89},
  number  = {4},
  pages   = {041003},
  year    = {2017},
  doi     = {10.1103/RevModPhys.89.041003}
}

@article{Kubo1954,
  title   = {Note on the stochastic theory of resonance absorption},
  author  = {Kubo, Ryogo},
  journal = {J. Phys. Soc. Jpn.},
  volume  = {9},
  number  = {6},
  pages   = {935--944},
  year    = {1954},
  doi     = {10.1143/JPSJ.9.935}
}

@incollection{Kubo1969,
  title     = {A stochastic theory of line shape},
  author    = {Kubo, Ryogo},
  booktitle = {Stochastic Processes in Chemical Physics},
  series    = {Advances in Chemical Physics},
  volume    = {15},
  pages     = {101--127},
  publisher = {Wiley},
  year      = {1969},
  doi       = {10.1002/9780470143605.ch6}
}

@article{KlauderAnderson1962,
  title   = {Spectral diffusion decay in spin resonance experiments},
  author  = {Klauder, J. R. and Anderson, P. W.},
  journal = {Phys. Rev.},
  volume  = {125},
  number  = {3},
  pages   = {912--932},
  year    = {1962},
  doi     = {10.1103/PhysRev.125.912}
}

@article{Krivnov1982,
  author  = {Krivnov, V. Ya. and Ovchinnikov, A. A.},
  title   = {Correlation functions of one-dimensional systems},
  journal = {Sov. Phys. JETP},
  volume  = {55},
  pages   = {162},
  year    = {1982},
  note    = {[Zh. Eksp. Teor. Fiz. \textbf{82}, 271 (1982)]}
}
\newpage

\renewcommand{\theequation}{S\arabic{equation}}
\renewcommand{\thefigure}{S\arabic{figure}}
\renewcommand{\thetable}{S\arabic{table}}

\setcounter{equation}{0}
\setcounter{figure}{0}
\setcounter{table}{0}
\setcounter{page}{1}
\renewcommand{\theequation}{S\arabic{equation}}
\renewcommand{\thefigure}{S\arabic{figure}}
\renewcommand{\thetable}{S\arabic{table}}
\renewcommand{\bibnumfmt}[1]{[S#1]}

\setcounter{equation}{0}
\setcounter{figure}{0}
\setcounter{table}{0}
\setcounter{page}{1}

\clearpage
\onecolumngrid
\title{Supplemental Material for \\
``Dephasing-Induced Melting of Crystalline Spectral Form Factors''}

\maketitle
\onecolumngrid
\vspace{-0.5cm}
\section{The reference DE Hamiltonian}

In the main text, we use the unfolded DE spectrum as our reference Hamiltonian. In this section, we show that the unfolding amounts to a simple spectral map. \\

As we know, the Dimitrui-Edelman (DE) construction provides a one-parameter family of real symmetric random matrices, where the Dyson index $\beta>0$ controls the eigenvalue repulsion. The eigenvalues fill out the Wigner semicircle of radius $R=\sqrt{2\beta d}$, with integrated density
\begin{equation}
\bar N(\lambda) = \frac{d}{\pi}\left[\arcsin\!\Big(\frac{\lambda}{R}\Big)+\frac{\lambda}{R}\sqrt{1-\Big(\frac{\lambda}{R}\Big)^{2}}\,\right]+\frac{d}{2}.
\end{equation}

Since $\bar N$ is a deterministic, monotonic function of a single real variable, we can apply it spectrally to $H_\beta$. Writing $H_\beta=U\Lambda U^{T}$, we define 
\begin{equation} 
H \equiv \bar N(H_\beta) = U\bar N(\Lambda)U^{T}, 
\label{eq:Hdef} 
\end{equation} 
which is real symmetric and commutes with $H_\beta$. The map is exact for any $d$ and $\beta$.  Its eigenvalues are
\begin{equation} 
x_n = \bar N(\lambda_n) = n + u_n, \qquad \langle u_n\rangle = 0, 
\end{equation}
with unit mean level spacing. 
This gives us the required near-crystalline energy spectra at large $\beta$, through a simple mapping.

\section{Spectral structure factor and bounds on the Bragg peaks}

Adopting the structure factor analogy, we have shown that the Bragg peaks, which are the coherent sum over the off-diagonal elements of the density matrix $\rho$, can be grouped via their energy gap labeled by $k=n-m$. We thus define the spectral structure factor
\begin{equation}
S_k(t)\ \equiv\ \sum_n \rho_{n,n-k}(t),
\label{eq:Sk}
\end{equation}
which measures the coherent, phase-sensitive amplitude carried by $\rho_{n,n-k}$. Given that $S_0=\mathrm{Tr}\,\rho=1$ and $S_{-k}=S_k^*$, the SFF is decomposed as~\cite{MartinezAzcona25}
\begin{equation}
d\,\mathrm{SFF}(t)=\sum_k S_k(t)=1+2\sum_{k>0}\mathrm{Re}\,S_k(t).
\label{eq:SFFfromSk}
\end{equation}

\begin{figure}[t]
  \centering
  \includegraphics[width=\textwidth]{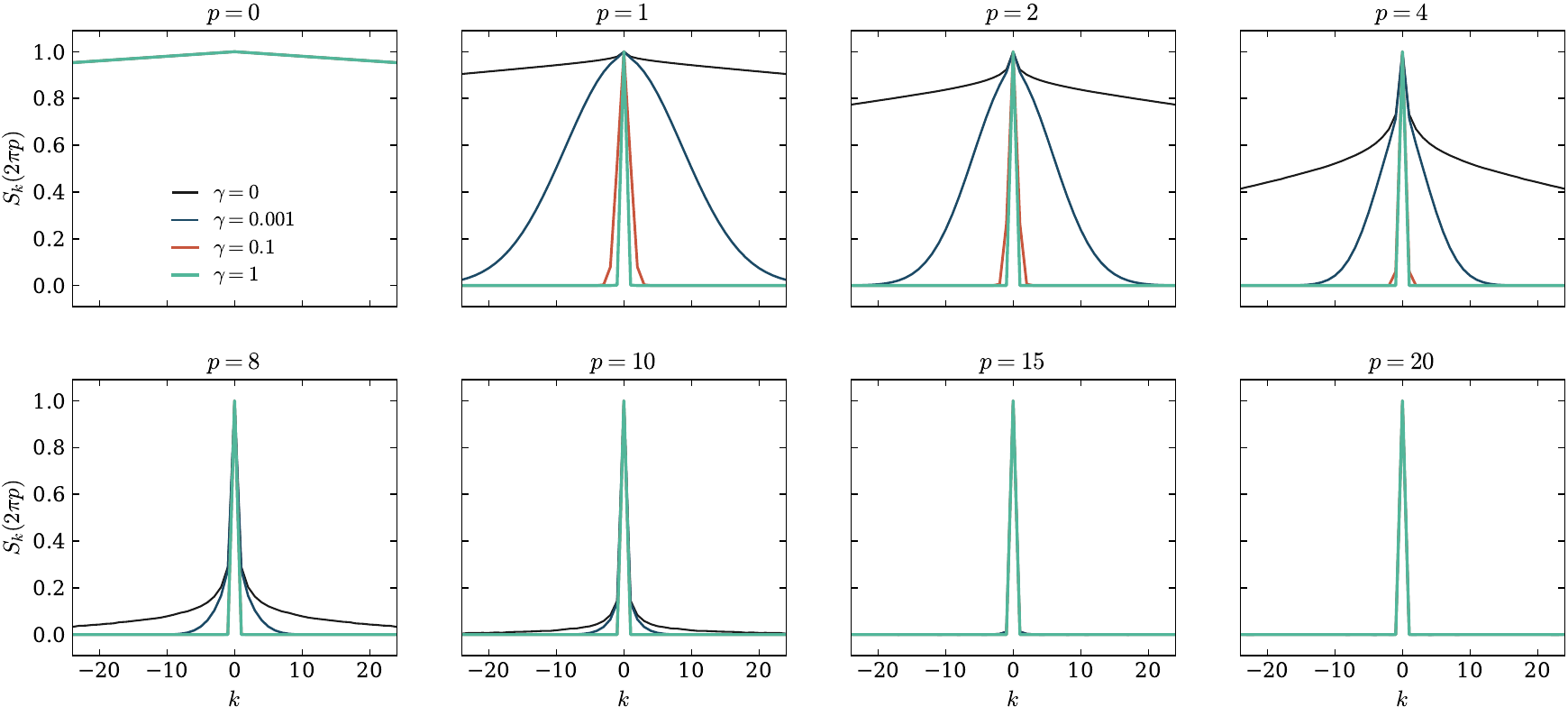}
  \caption{%
    Melting of the crystalline spectral order as the narrowing of the spectral structure factor.
    Ensemble-averaged $S_k(2\pi p)=\sum_n\rho_{n,n-k}$ across Bragg orders
    $p=0,1,2,4,8,10,15,20$, for $\gamma\in\{0,10^{-3},10^{-1},1\}$. At $p=0$ all channels coincide with the reference triangle $S_k(0)=1-|k|/d$, a finite-size multiplicity giving $K(0)=\frac1d\sum_k S_k(0)=1$. Each subsequent panel is this triangle multiplied by the two Debye--Waller factors $W_p^\beta$ and $W_p^\gamma$. Energy dephasing contributes $e^{-2\pi\gamma p k^2}$, which removes the higher orders first and narrows the profile with increasing $\gamma$ at fixed $p$. Intrinsic level fluctuations lead to the term $e^{-\frac12(2\pi p)^2g(k)}$, which disperses the phases and lowers all orders, so that even the unitary ($\gamma=0$) profile narrows as $p$ grows. The Bragg peak height $K_0(2\pi p)=\frac1d\sum_kS_k$ is set by the width of each profile, and reaches the plateau $1/d$ once only $k=0$ survives.
  }
  \label{fig:Sk_grid}
\end{figure}

In this picture, the decay of the Bragg peak heights can be seen as the narrowing of the effective width of $S_k$. Under energy dephasing evolution, $S_k$ acquires two independent Debye--Waller like suppressing terms,
\begin{equation}
S_k(t_p)=\Big(1-\tfrac{|k|}{d}\Big)\,
e^{-2\pi\gamma p\,k^{2}}\,
e^{-\frac{1}{2}(2\pi p)^{2}g(k)},
\label{eq:Sk_dw}
\end{equation}
where $g(k)=\mathrm{Var}(\delta_n-\delta_{n-k})$.
The first exponential is a dynamic factor: decoherence at rate $\gamma$ suppresses the outer diagonals, with a cutoff $k_\ast=(2\pi\gamma p)^{-1/2}$ that shrinks with $p$. The second is a static factor set by intrinsic level fluctuations, which suppresses all diagonals quadratically in $p$ and persists even at $\gamma=0$~\cite{TruninHuse2026}. Figure~\ref{fig:Sk_grid} illustrates $S_k$, for different $\gamma$, at successive Bragg times, starting from $p=0$. The profile narrows with $p$ even in the unitary case, and increasingly so as $\gamma$ grows. \\

Here, we further show that the spectral structure factor offers a tighter bound for the crystalline spectral form factor. 

The upper and lower bounds of the SFF in terms of the  $\ell_1$-norm read ~\cite{BaumgratzCramerPlenio2014,
MatsoukasRoubeas2024},
\begin{equation}
\frac{1-C_{l_1}(t)}{d}\ \le\ \mathrm{SFF}(t)\ \le\
\frac{1+C_{l_1}(t)}{d},
\end{equation}
where the coherence monotone $C_{l_1} = \sum|\rho_{n,n-k}|$ is insensitive to phase misalignment. The values $|\rho_{nm}|=1/d$ are conserved, and $C_{l_1}=d-1$ remains maximal in the unitary evolution, even though the Bragg peaks still collapse due to the phase scrambling induced by the fluctuations $\delta_n$. The structure factor corrects this. Applying the triangle inequality to Eq.~\eqref{eq:Sk} and defining $\Sigma(t)\equiv\sum_{k\neq0}|S_k(t)|$, Eq.~\eqref{eq:SFFfromSk} gives the tighter bound
\begin{equation}
\frac{1-\Sigma(t)}{d}\ \le\ \mathrm{SFF}(t)\ \le\
\frac{1+\Sigma(t)}{d},
\qquad \Sigma(t)\ \le\ C_{l_1}(t).
\label{eq:Skbound}
\end{equation}
The difference $C_{l_1}-\Sigma =\sum_{k\neq0}\!\big(\sum_n|\rho_{n,n-k}|
-|\sum_n\rho_{n,n-k}|\big)\ge0$ is the alignment deficit. It accounts for coherences that are populated but not phase-locked. As one would expect, for perfect crystalline order, this quantity vanishes at Bragg times and stays strictly positive once the fluctuations $\delta_n$ are introduced.

\section{Order-Resolved Decay of Neighbor Coherences}

In this section, we examine in detail how intrinsic fluctuations and extrinsic energy dephasing affect the neighbor-order coherences. We recall that the SFF can be decomposed into a sum of $k$-th neighbor coherences, which together carry the entire structure of the SFF. \\

\begin{figure}[h]
    \centering
    \includegraphics[width=0.80\columnwidth]{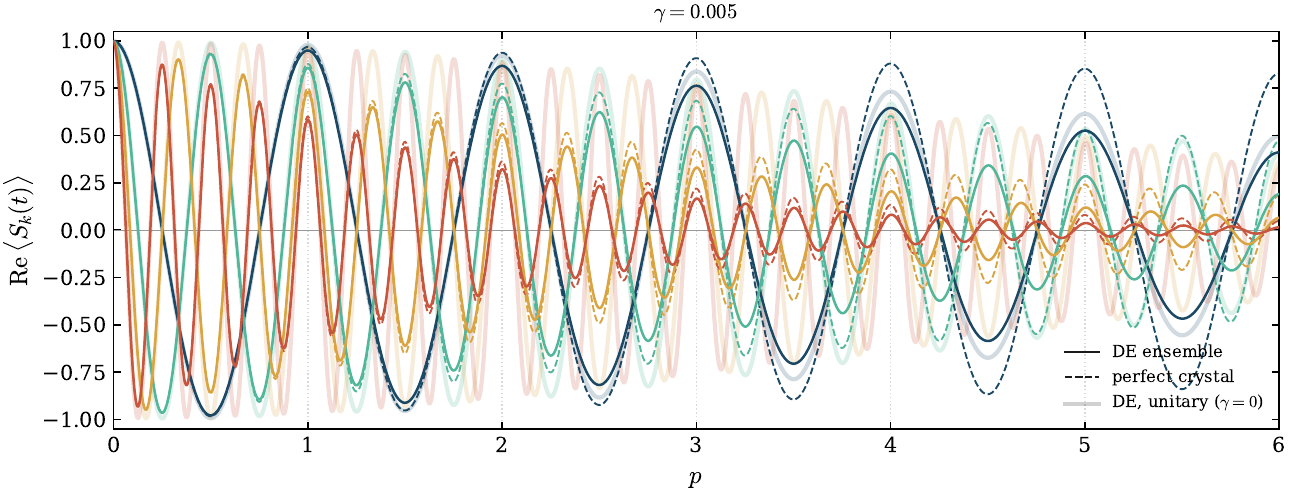}
    \includegraphics[width=0.80\columnwidth]{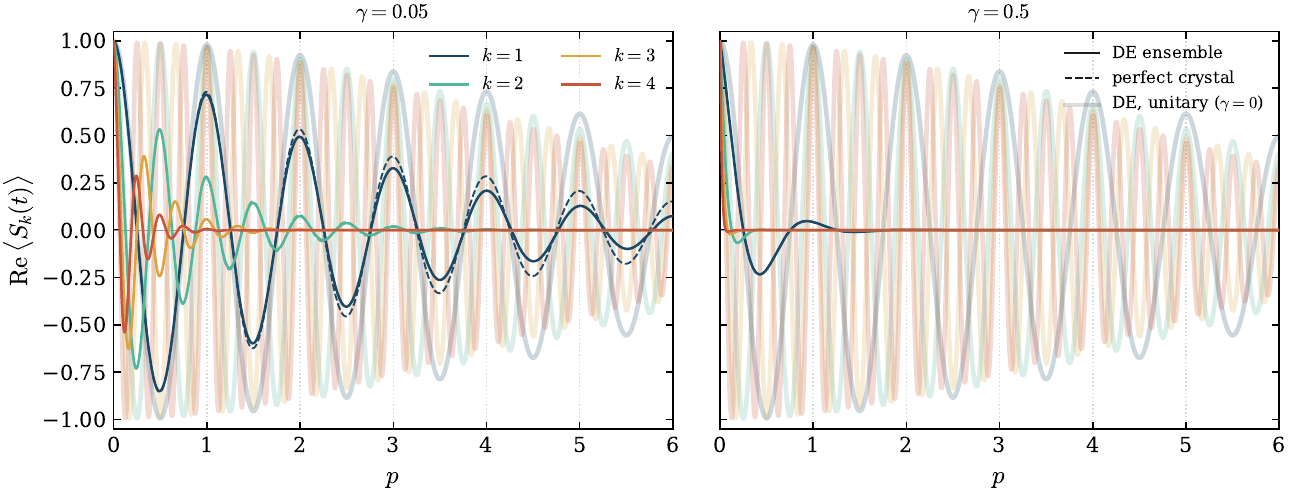}
    \caption{Neighbor-order contributions $\mathrm{Re}\langle S_k(t)\rangle$, $k=1$--$4$, in the unfolded Dumitriu--Edelman ensemble ($\beta=500$, $d=512$, solid) and the perfect crystal (dashed), at (a) $\gamma=0.005$ and (b) $\gamma=0.1$. All orders align at integer $p=t/t_H$. The Background shows $\mathrm{Re}\langle S_k(t)\rangle$ for unitary evolution, in the absence of dephasing.}
    \label{fig:Sk}
\end{figure}

Figure~\ref{fig:Sk} shows the alignments of individual neighbor-order coherences at integer $t/t_H$. By comparing the DE ensembles with the perfect crystalline spectrum, we isolate the effect of intrinsic fluctuations: the amplitudes decrease, but the oscillations remain exactly in place. Increasing $\gamma$, rapidly reduces the higher-order contributions, as seen in Fig.~\ref{fig:Sk}(b), for $\gamma=0.1$. The peak heights are then dominated by energy dephasing rather than intrinsic fluctuations, as shown by comparing the DE ensemble peaks under energy dephasing with unitary evolution. Between revivals, each order oscillates with its own period $t_H/k$, so only subsets of orders align, and full Bragg peaks appear only at integer $t/t_H$. The amplitude also decreases with $k$, since the relative displacements $g(k)$ increase with the gap.

\section{Intrinsic Debye--Waller factor and spectroscopy of spectral rigidity}

In the crystalline spectral form factor, energy dephasing not only induces decoherence but also serves as a useful tool to probe spectral rigidity. In this section, we show that by sweeping $\gamma$, one can identify the type of spectral order.\\

We recall that the intrinsic fluctuations show up in the Bragg peaks as
\begin{equation}
C_p(k)=\Big|\,\langle e^{-it_p(x_n-x_{n-k})}\rangle \,\Big|,\qquad t_p=2\pi p,
\label{eq:SM_Cp}
\end{equation}
where the average over different spectral realizations is denoted by $\langle\cdot\rangle$. It measures the phase alignment within the neighbor-order $k$ at the $p$-th revival, and coincides with $S_k(t_p)/S_k(0)$ in the absence of dephasing. For Gaussian displacements, $C_p(k) = e^{-\frac12t_p^2g(k)}$, with $g(k)=\langle(u_n-u_{n-k})^2\rangle$. Under energy dephasing, the coherent distribution to the Bragg peak gets modified as
\begin{equation}
e^{-2W}=\frac2d\sum_{k=1}^{d-1}\Big(1-\frac kd\Big)\,
e^{-k^2/2k_\ast^2}\,C_p(k),\qquad k_\ast=(4\pi\gamma p)^{-1/2}.
\label{eq:SM_DW}
\end{equation}
Here, decoherence restricts the sum to the gap $k\lesssim k_\ast$, while $C_p(k)$ weights the neighbor-orders within this window.\\

\begin{figure*}
\centering
\includegraphics[width=0.85\columnwidth]{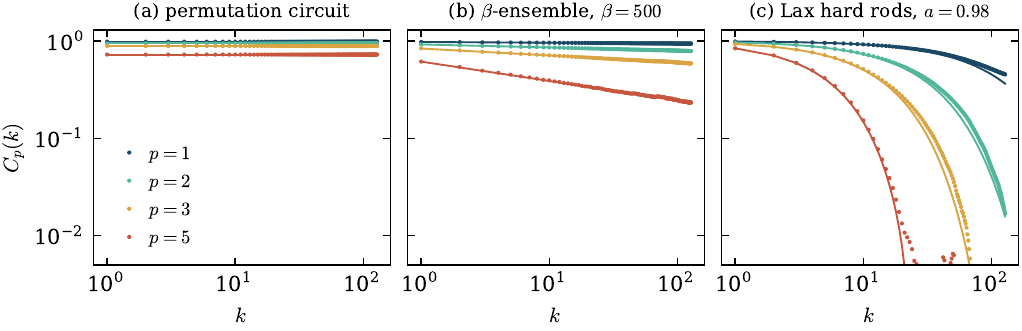}
\caption{Intrinsic Debye--Waller factor $C_p(k)$, Eq.~\eqref{eq:SM_Cp}, at revival orders $p=1,2,3,5$ for (a) a permutation-symmetric circuit (LRO), (b) the $\beta$-ensemble with $\beta=500$ (QLRO), and (c) Lax hard rods with $a=0.98$ (SRO). Symbols show numerical data, and lines the Gaussian prediction $e^{-\frac12t_p^2g(k)}$ with the measured $g(k)$.}
\label{fig:SM_Cp}
\end{figure*}

We first compute $C_p(k)$ for three models with qualitatively different spectral order and later show that it can be identified using the slope of the Bragg-peak response to $\gamma$. Building on the models analyzed under unitary dynamics in Ref.~\cite{TruninHuse2026}, Fig.~\ref{fig:SM_Cp} compares three classes of spectral order. 

In a permutation-symmetric circuit, $g(k)$ remains bounded and $C_p(k)$ is independent of $k$, reflecting long-range spectral order (LRO). The suppression then reduces to a gap-independent factor that decreases with $p$. 

Following the example used in the main text, the $\beta$-xxensemble has logarithmic correlations of the Dyson gas, $g(k)\simeq g_1+\frac{c}{\beta\pi^2}\ln k$, and $C_p(k)$ results in an algebraic decay $C_p(k)\propto k^{-\eta_p}$ with $\eta_p=2cp^2/\beta$, which resembles the characteristic feature of quasi-long-range order (QLRO). From the numerical data in Fig.~\ref{fig:SM_Cp}(b), we get $c\simeq2$. 

Lastly, the Lax hard rods~\cite{Bogomolny2009Lax,Bogomolny2011} are described by levels that behave as rods of length $a$ in units of the mean level spacing. Each spacing consists of the rod length plus a fluctuating free gap of mean $1-a$, and the perfect crystal is recovered in the limit $a\to1$. The spacings fluctuate independently about their mean, so the displacement of a level relative to one $k$ sites away is the sum of $k$ independent spacing errors, and it grows like a random walk, $g(k)=\sigma_s^2k$, with $\sigma_s\simeq1-a$ as the spread of a single spacing. The phase misalignment therefore accumulates linearly with the gap, and $C_p(k)\simeq e^{-k/\xi_p}$ decays exponentially beyond a scale $\xi_p$ that decreases with $p$, as expected in short-range order (SRO).\\

Since the dephasing window samples $C_p(k)$ at the scale $k_\ast$, each class leaves distinct signatures in the dependence of the Bragg peaks on $\gamma$. Specifically, differentiating Eq.~\eqref{eq:SM_DW} with respect to $\ln\gamma$ isolates gaps at $k\simeq\sqrt2\,k_\ast$, and the exponent $\tilde\alpha_p\equiv\partial\ln|\partial K(t_p)/\partial\ln\gamma|/\partial\ln\gamma$ tracks the local slope of $C_p(k)$ at this scale. For $1\ll k_\ast\ll d$, this yields
\begin{equation}
\frac{\partial\ln\big|\partial K(t_p)/\partial\ln\gamma\big|}{\partial\ln\gamma}
\equiv\tilde\alpha_p=
\begin{cases}
-\tfrac12, & \text{LRO},\\[2pt]
-\tfrac12+\tfrac{\eta_p}{2}, & \text{QLRO},\\[2pt]
-\tfrac12\to1, & \text{SRO}\ (k_\ast\sim\xi_p).
\end{cases}
\label{eq:alpha}
\end{equation}

\begin{figure}[b]
    \centering
    \includegraphics[width=0.65\columnwidth]{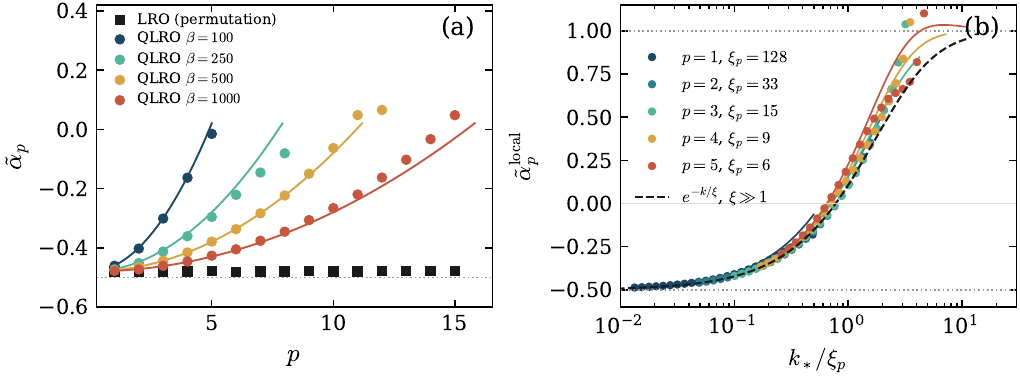}
    \caption{Spectroscopy of spectral order. (a) Response exponent $\tilde\alpha_p$, Eq.~\eqref{eq:alpha}, versus revival order $p$ for a permutation-symmetric circuit (LRO, squares) and the $\beta$-ensemble (QLRO, circles) at $\beta=100$, $250$, $500$, and $1000$. Lines show the prediction $\tilde\alpha_p=-\tfrac12+\eta_p/2$ with $\eta_p=2cp^2/\beta$; the dotted line marks the crystalline value $-\tfrac12$. (b) Response exponent for Lax hard rods (SRO) versus $k_\ast/\xi_p$ at $p=1$--$5$. Solid lines show Eq.~\eqref{eq:SM_DW} with $C_p(k)=e^{-k/\xi_p}$, and the dashed line its limit $\xi_p\gg1$. Dotted lines mark the limiting values $-\tfrac12$ and $1$.}
    \label{fig:alpha_p}
\end{figure}

In Fig.~\ref{fig:alpha_p}, we sweep $\gamma$ and identify the exponent $ \tilde\alpha_p$ from the numerical data, finding good agreement with the analytical prediction. The $\gamma$-independent prefactor from the $k-$independent part of $g(k)$ drops out of the exponent. Consequently, the exponent characterizes the growth of $g(k)$ with the gap. Tuning the dephasing strength therefore enables a form of spectroscopy for spectral rigidity. In the case of $\beta$-ensembles, $\tilde\alpha_p$ changes sign at $\eta_p=1$, near the critical revival order $p_c$.

\end{document}